\documentstyle[11pt,newpasp,twoside,psfig]{article}
\def\ltsima{$\; \buildrel < \over \sim \;$}
\def\simlt{\lower.5ex\hbox{\ltsima}}            
\def\gtsima{$\; \buildrel > \over \sim \;$}
\def\simgt{\lower.5ex\hbox{\gtsima}}            

\begin{document}

\title{The AGN Paradigm for Radio-Loud Objects}
\author{Meg Urry}
\affil{Yale Center for Astronomy and Astrophysics,
Yale University, PO Box 208121, New Haven CT 06520-8121, USA}

\begin{abstract}
Radio-loud AGN are characterized by relativistic jets
originating near the central supermassive black hole
and forming large-scale radio sources at parsec to 
kiloparsec to Megaparsec distances. The jets are energetically
significant, in many cases representing the bulk of the energy
extracted from the accretion process.
Host galaxies are apparently normal luminous ellipticals, 
supporting the ``Grand Unification'' scenario wherein 
AGN are a transient phase in the evolution of every galaxy. 
Black hole mass appears to be largely uncorrelated
with bolometric luminosity, Eddington ratio,
radio luminosity, or radio loudness.
\end{abstract}

\section{Accepted Wisdom}

According to the paradigm for active galactic nuclei (AGN), illustrated in Figure~1, 
gravitational potential energy is converted to radiation and
kinetic power via accretion of matter onto a central supermassive
black hole. We observe thermal emission from the disk in the 
optical, ultraviolet, and X-ray; 
Compton scattered X-rays from a hot corona;
Thomson scattering of the nuclear continuum in an extended corona 
(Antonucci \& Miller 1985); 
and broad and narrow emission lines.
From some directions, the nuclear continuum and broad lines are obscured 
by a torus or warped disk; whether the torus geometry changes with 
luminosity and/or redshift will be settled by upcoming
SIRTF searches for highly obscured quasars at redshift $z\sim2-3$.
These characteristics of AGN are independent of radio loudness.

\begin{figure}[t]
\centerline{
\psfig{figure=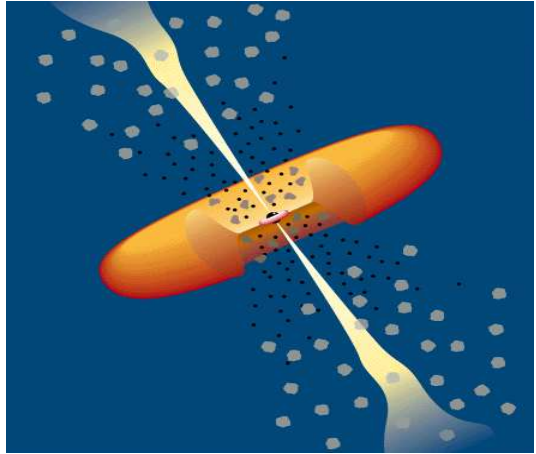,width=7cm,height=6.0cm,angle=0}
}
\caption{Schematic drawing of the key elements of an active galactic nucleus:
central supermassive black hole, accretion
disk, broad-line region surrounded by a dusty torus or warped disk,
extended corona of hot electrons,
extended narrow-line region, and, at least for
radio-loud AGN, a pair of relativistically outflowing jets which
originate near the black hole and eventually decelerate on large scales.  }
\label{fig1}
\end{figure}

The distinguishing feature of radio-loud AGN is a pair
of relativistic jets, originating within a few tens of Schwarzschild
radii of the black hole. 
The relativistic potential well of a black hole --- whether a supermassive
black hole at the center of a distant galaxy or a few-solar-mass black hole
in our Galaxy --- appears to lead naturally to relativistic
jet speeds.
Weaker jets disrupt sooner and thus have a diffuse appearance
(these are Fanaroff-Riley type 1 sources),
while the most powerful jets remain well collimated to large distances
from the nucleus, where they form large double
radio lobes with hot spots (Fanaroff-Riley type 2 sources).
It is possible that radio-quiet AGN have such weak jets that they
are quenched before developing into observable extended radio sources.

Probably the black hole spin (or accretion disk rotation) defines 
an axis of symmetry along which material is ejected to form the jets.
In some cases, this axis must be stable over very long times, allowing
the jets to extend to a Megaparsec or more (e.g., NGC~6251). 
In other cases, precession of the black hole, or its motion through a dense
environment, are detectable from the distorted shapes of the jets
(e.g., Wide-Angle Tail sources).
AGN with jets viewed end-on are classified as blazars 
(either BL Lac objects or Optically Violently Variable quasars), 
while ordinary quasars are viewed at larger angles from the
jet axis, and radio galaxies have jets more nearly in the plane
of the sky.

Jets are relativistic on parsec scales, as attested 
by the ubiquity of superluminal motion in blazars. Systematic 
VLBI (\& VSOP) surveys of core-dominated radio sources 
confirm this:
in addition to bright cores, these radio sources have
high optical polarization,
jet pc-kpc misalignments,
one-sided morphologies, and
intraday variability (in the radio), 
all characteristics explained naturally by relativistic beaming
(see Lister et al., this conference). 

Jets are very likely relativistic on kiloparsec scales as well.
Large-scale relativistic proper motions have been directly observed
in the nearby radio galaxy M~87 (Biretta et al. 1995).
Jet one-sidedness is common (Bridle \& Perley 1984),
and correlates well with VLBI one-sidedness (which is 
surely due to relativistic beaming)
and with depolarization asymmetry (Laing 1988; Garrington and Conway 1991). 
More recently, the most plausible explanation of some 
of the newly discovered extended X-ray jets requires that
the plasma have bulk relativistic motions on scales of hundreds
of kiloparsecs (Tavecchio et al. 2001, Celotti et al. 2001,
Sambruna et al. 2002).

In powerful radio sources, jets are straight and remain relativistic to large
distances (a well-known example is the radio galaxy Cygnus~A).
In less powerful radio sources (for example, M~87), 
jets become sub-relativistic closer to the
radio core, but they likely are still relativistic initially 
(Giovannini et al. 1998).
There are systematic differences in the emission line properties
of AGN as a function of radio power:
low-luminosity (FR~1) sources have low excitation lines,
while high-luminosity (FR~2) sources are more diverse, 
having a range of ionization states and line strengths
(Laing et al. 1994; Chiaberge et al. 2002 and this conference).
Clearly, some unify (through orientation) with luminous BL Lacs, 
while others are quasars. 
Interestingly,
there appears to be an analogous situation among radio-quiet objects, 
in that some Seyfert~2s have hidden (strong) broad-line regions
while others simply have intrinsically weak lines (Tran 2001).

Population statistics confirm that relativistic beaming
is an essential feature of radio-loud AGN.
Luminosity functions and number counts for blazars are
commensurate with their being beaming-dominated, 
aligned versions of normal radio galaxies (Urry \& Padovani 1995).
The bulk Lorentz factors are $\Gamma\sim 5-10$,
similar to those inferred from superluminal motion. 
Some have speculated that OVV quasars might evolve into BL Lac
objects, which is equivalent to saying that powerful
FR~2 radio sources evolve into FR~1 radio sources
(Vagnetti et al. 1991, Maraschi \& Rovetti 1994,
Jackson \& Wall 1999).

\section{Underlying Physics and Jet Energy}

\begin{figure}[t]
\centerline{
\psfig{figure=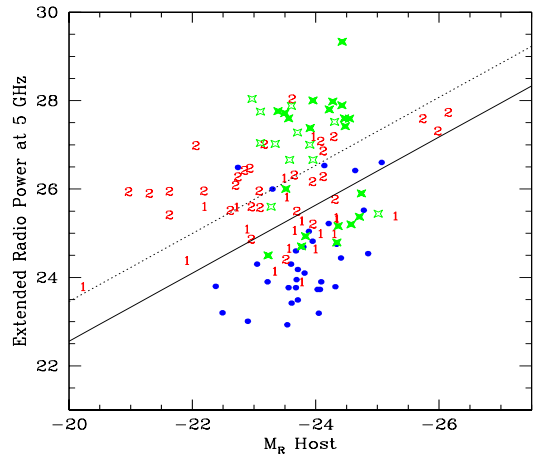,width=7cm,height=6cm}
}
\caption{Radio power versus host galaxy magnitude for radio sources.
The diagonal division between low-luminosity sources 
(FR~1 galaxies and BL Lac objects) and
high-luminosity sources (FR~2 galaxies and quasars) 
suggests that jets with higher power can punch through
galaxies with greater mass (greater interstellar density).
(Figure from Urry et al. 1999, following Owen \& Ledlow 1994.)
 }
\label{fig2}
\end{figure}

Orientation is obviously a major influence on observed properties of 
radio-loud AGN. The jet brightness can vary by factors of a million or 
more, depending on whether it is approaching, receding,
or in the plane of the sky.
Yet intrinsic physical differences among AGN are clearly present, and
are the interesting part of the story. How do we get at these
intrinsic AGN characteristics? 
Probably the most important are mass accretion rate, 
efficiency of converting gravitational potential energy to radiation 
and kinetic outflows, and black hole mass. 
A combination of the first two can be guessed from thermal emission from the
accretion disk (or equivalent) -- UV bump, lines, etc.

In evaluating the extraction of energy from accretion, however,
one needs to account not only for the observed radiation
but for the power channeled into the jet, which is
considerable (Celotti \& Fabian 1993). 
In many cases, the kinetic power dominates the total power budget.

Jets must be produced with a range of power, possibly even
a bimodal distribution (Meier 2001).
Observationally, however, the distribution of jet powers 
is not really known. It is an important constraint on models 
for jet formation, and an essential ingredient in models
that explain the large-scale morphologies of radio galaxies. 
It may be possible to infer the distribution of jet powers
from studies of blazar jets, and a number of such studies
are underway. 

There is clear evidence (Fig.~2) that the morphologies of radio sources 
depend on both intrinsic jet power and on the density of
surrounding medium (Owen \& Ledlow 1994), and indeed
models have been successfully developed along these lines
(Bicknell 1985, De Young 1993).

\section{Host Galaxies, AGN Formation, and Black Hole Growth}

The most powerful radio galaxies formed at high redshifts, 
judging from the colors of their host galaxies. 
They are similar to the host galaxies of nearby,
less powerful radio galaxies, apart from passive stellar evolution.
Ultraviolet images of local radio galaxies are very 
similar to the rest-frame ultraviolet images of distant radio galaxies,
apart from their lower luminosity.
Probably radio-loud AGN have more luminous host galaxies
than radio-quiet objects.

There is evidence that the onset of radio activity --- 
namely, the formation of jets --- is associated
with the clearing of dense gas and dust,
based on associated CIV absorption (Baker et al. 2002). 
The frequent association of AGN and starbursts 
is also commensurate with an emerging picture
wherein the onset of accretion (and thus extraction of energy 
from the central black hole) coincides with formation of 
the galactic bulge (a dissipative collapse, possibly merger induced)
and with the ensuing starburst activity. 
These events may be truly simultaneous in cosmic terms 
(Kormendy \& Gebhardt 2000).

\section{Radio-Quiet and Radio-Loud AGN}

It is entirely possible that relativistic jets are produced 
in the centers of all AGN, including radio-quiet objects;
in the latter sources, they simply become
sub-relativistic much sooner. 
There is some tentative evidence for this.
There are suggestions that relativistic beaming is
important in intermediate radio-loudness objects
(namely, those with flat-spectrum radio cores; Falcke et al. 1996),
and superluminal motion has been reported 
in two radio-quiet objects (Blundell 1998, Brunthaler et al. 2000).

\begin{figure}[t]
\centerline{
\psfig{figure=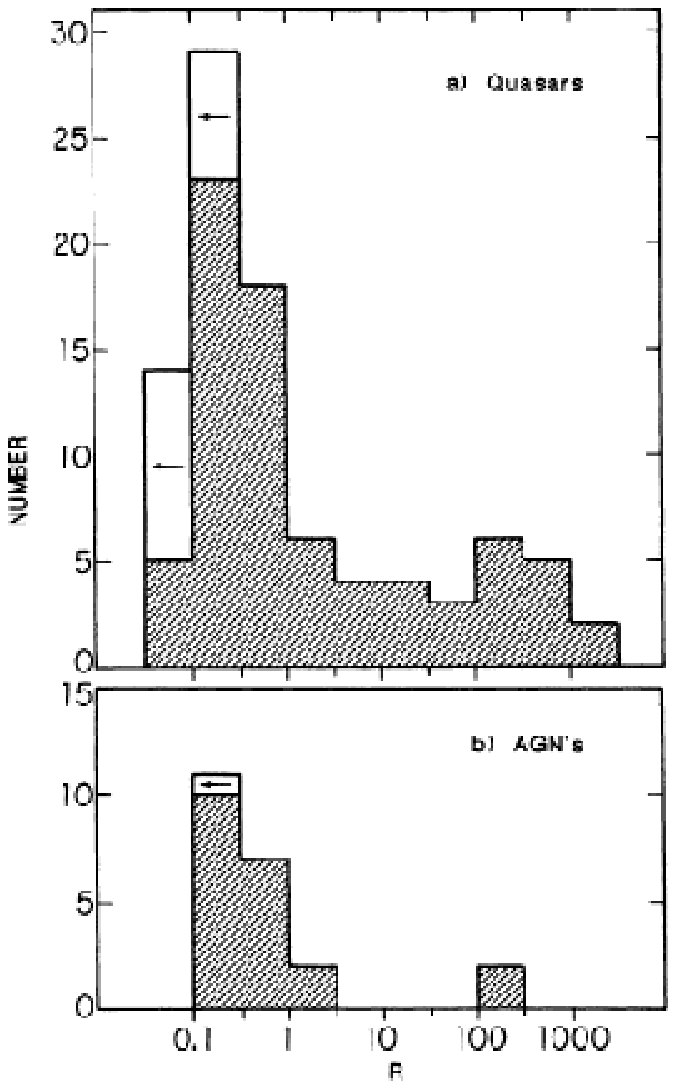,width=4.5cm,height=6.2cm,angle=0}
\psfig{figure=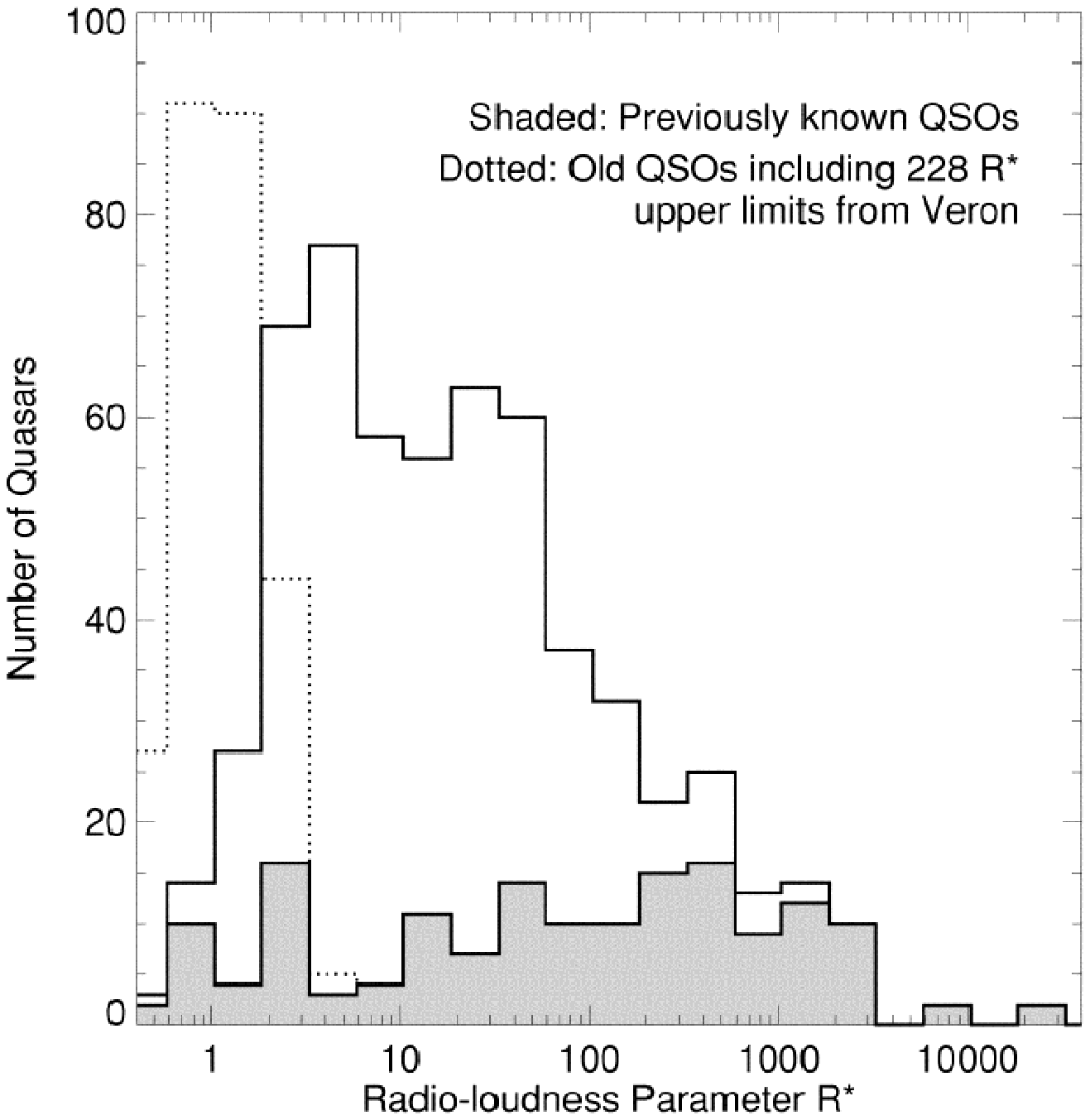,width=6.0cm,height=7.0cm,angle=0}
}
\caption{Distributions of radio loudness for two quasar samples.
{\it Left:} PG quasars (Kellerman et al. 1989);
{\it right:} FIRST quasars (White et al. 2001).
These distributions are not convincing evidence of bimodality in 
radio loudness; indeed, the FIRST distribution suggests
a continuous distribution of radio loudness, and thus
no sharp division between radio-loud and radio-quiet AGN. }
\label{fig3}
\end{figure}

The notion of a distinct bimodality in radio loudness of AGN ---
that is, the separate categories ``radio-quiet'' and ``radio-loud'' ---
is based largely on the radio properties of the PG sample
of quasars (Kellerman et al. 1989). But the statistics of that
sample were
not overwhelming and indeed the data are compatible with
a roughly Gaussian peak at low radio-to-optical ratio 
(representing the 90\% of AGN that are radio-quiet)
plus an extended tail to higher values 
(the 10\% of AGN that are radio-loud).
The FIRST bright quasar sample, which probes much lower radio fluxes,
indeed shows no signs of a bimodal distribution in radio
loudness (White et al. 2001).
Figure~3 shows the two distributions.\footnote{We 
use the standard definition of
radio-loudness, $R>10$, where $R \equiv F_{5~{\rm GHz}} / F_{B}$
(Kellerman et al. 1989).}

\begin{figure}[t]
\centerline{
\psfig{figure=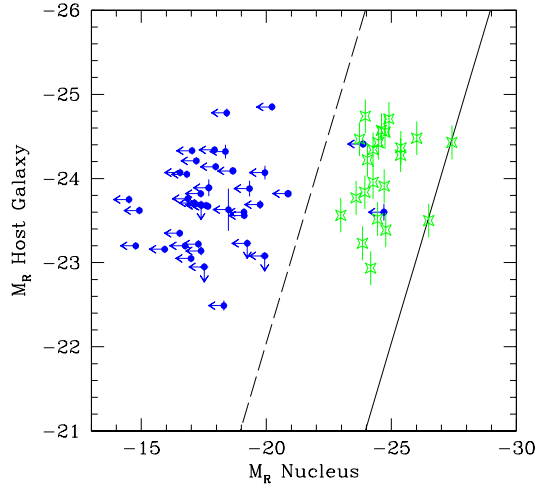,width=7cm,angle=0}
}
\caption{Absolute R-band host galaxy versus nuclear magnitude 
(K-corrected and, for BL Lac nuclei, also corrected for beaming) 
for low-power radio-loud AGN ({\em filled circles}) 
and high-power radio-loud AGN ({\em stars}).
The scatter in the host galaxy
magnitudes is very small (rms $\sim0.25$ in $\log L_{gal}$), 
over a range of more than 4 orders
of magnitude in nuclear luminosity. 
The {\em solid line} indicates
an Eddington ratio of $L/L_{Edd}=1$, and the
{\em dashed line} is $L/L_{Edd}=0.01$, obtained by assuming the
host galaxy luminosity--black hole mass relation reported by
Kormendy \& Gebhardt (2001). }
\label{fig4}
\end{figure}

\section{Black Hole Masses}

Black hole mass is a fundamental property of AGN,
and must be related, in an as-yet unknown way, to the
individual characteristics of AGN. In recent years there has been
a flurry of activity on this front.

Black hole mass can be estimated in several ways.
Direct dynamical methods --- for example, measuring orbital
motions of gas very near the black hole --- are possibly only for
a few very local AGN (e.g., NGC4258; Miyoshi et al. 1995).
More generally, one can apply the virial theorem to
the broad emission line clouds, provided the distance from the
black hole to the broad line region is known. This distance can be
estimated most reliably from reverberation mapping, but can
also be guessed from photoionization models, or from
optical or ultraviolet luminosity (which appears to correlate
with reverberation-mapped broad-line region size; Kaspi et al. 2000).

\begin{figure}[t]
\centerline{
\psfig{figure=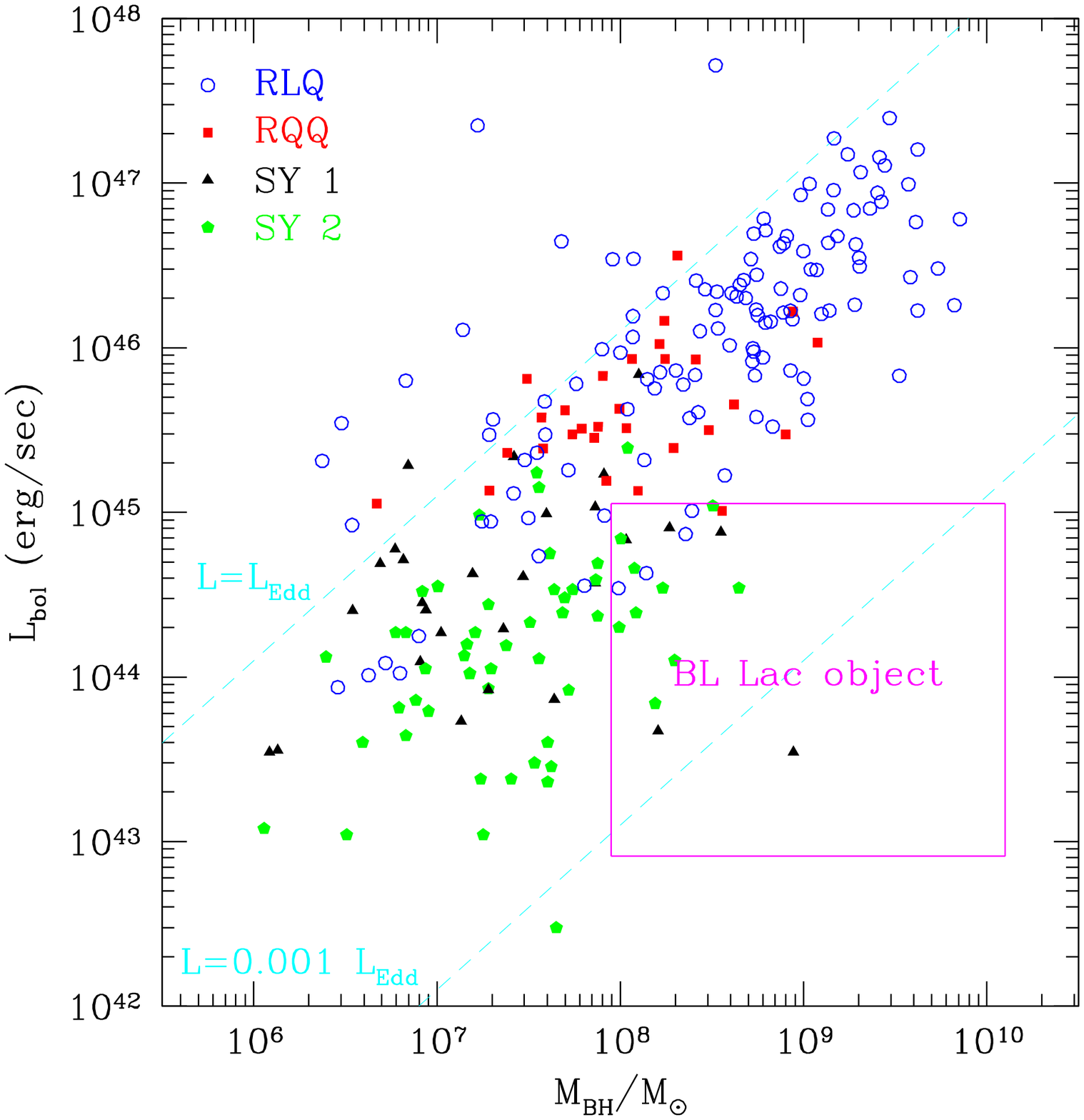,width=5.5cm,height=5.8cm,angle=0}
\psfig{figure=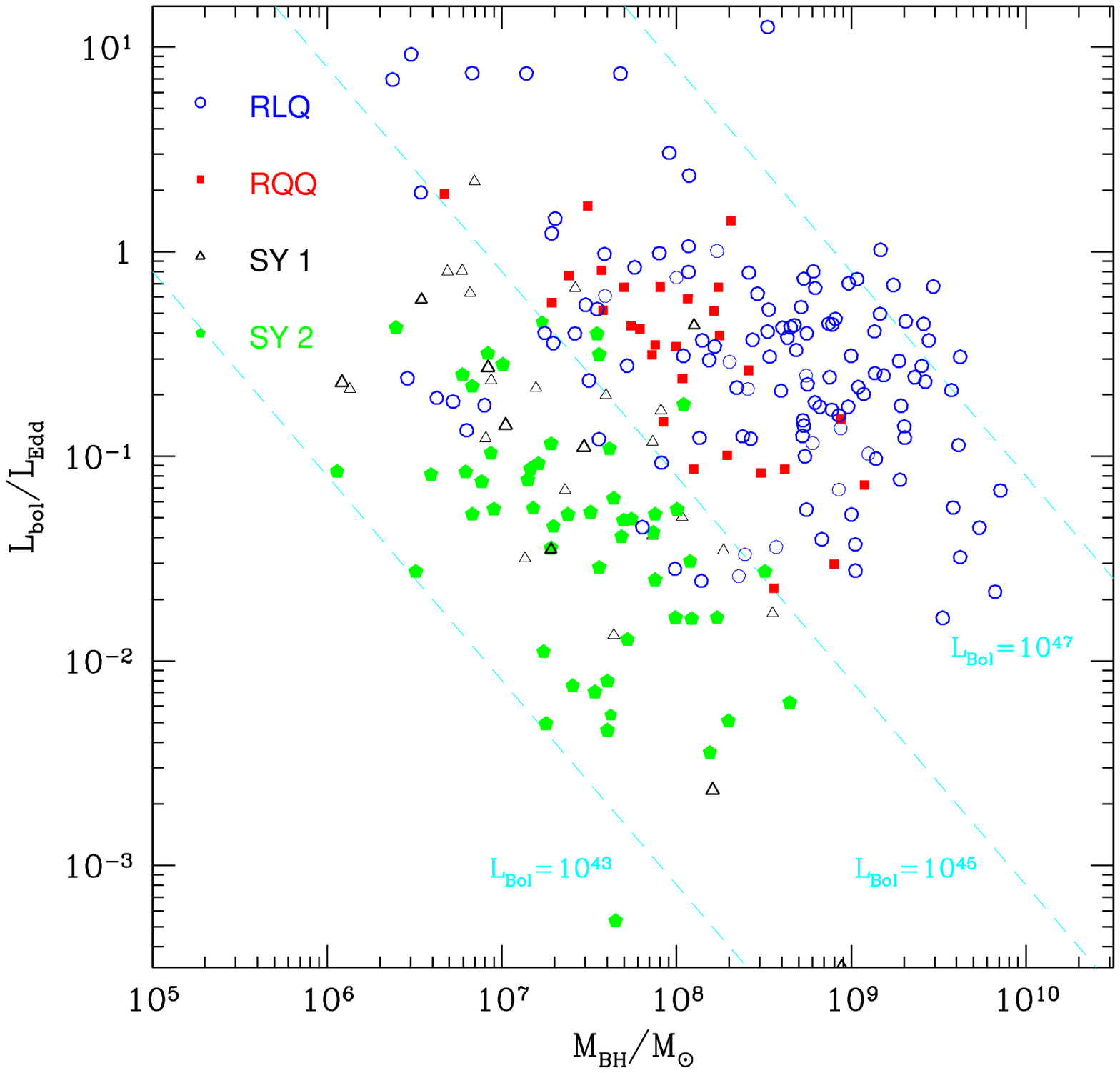,width=6cm,height=6cm,angle=0}
}
\caption{Bolometric luminosity ({\it left}) and Eddington ratio ({\it right})
versus black hole mass.
There is little if any correlation. 
{\it Left:} The Eddington limit gives an approximate upper limit 
({\it upper diagonal dashed line}) to the observed luminosity
but the rest of the $L_{bol}$ vs. $M_{BH}$ space is filled.
(Objects with significant relativistic beaming
or orientation-dependent obscuration are not included in this figure;
the box shows where BL Lac objects (after beaming correction) would like
in the lefthand figure.)
{\it Right:} Eddington ratios span 2--4 orders of
magnitude at a given black hole mass.
Unpopulated regions of this diagram are due to
selection effects (see Woo \& Urry 2002a).}
\label{fig5}
\end{figure}

Alternatively, we can infer black hole mass from 
host galaxy properties,
under the assumption that AGN host galaxies
are intrinsically the same as non-active galaxies
(the ``Grand Unification'' hypothesis, 
for which there is increasing evidence). 
Normal galaxies show a clear correlation between black hole mass
and stellar velocity dispersion, $\sigma$, in the galaxy. 
Thus if one could measure $\sigma$ in AGN hosts, or infer it somehow
(for example, from the morphological parameters $\mu_e$ and $r_e$,
together with the fundamental plane correlation for early type galaxies),
then the $M_{BH} - \sigma$ correlation yields a black hole mass estimate.

Our extensive HST survey of the host galaxies of low-luminosity 
radio-loud AGN 
(Falomo et al. 1997, 2000; Urry et al. 1999, 2000;
Scarpa et al. 2000, 2001)
showed that the host galaxy properties were remarkably independent
of nuclear luminosity.
Essentially all hosts are luminous elliptical galaxies,
which follow well the Kormendy relation between half-light
radius and surface brightness at that radius 
(the $\mu_e - r_e$ anti-correlation).
Comparing our sample to host galaxies of higher luminosity radio-loud
AGN (quasars), we found that galaxy luminosity was remarkably 
uniform, despite several orders of magnitude range in nuclear luminosity
(O'Dowd et al. 2002; Fig.~4).

To extend this study to more AGN, both radio-loud and radio-quiet,
we collected black hole mass estimates for a heterogeneous sample
of nearly 500 AGN (Woo \& Urry 2002a,b), using both the virial method 
and the host-galaxy method. 
We used published multiwavelength data to estimate bolometric luminosities 
for every AGN in the sample.
The goal was to look for trends of black hole mass with 
luminosity, Eddington ratio, radio loudness, and so on.
We found no significant trends (or rather, none that could not be 
explained by obvious selection effects like flux limits or volume surveyed).
As implied by the earlier O'Dowd work, black hole mass is not correlated
with luminosity (Fig.~5a) or with Eddington ratio (Fig.~5b).

\begin{figure}[t]
\centerline{
\psfig{figure=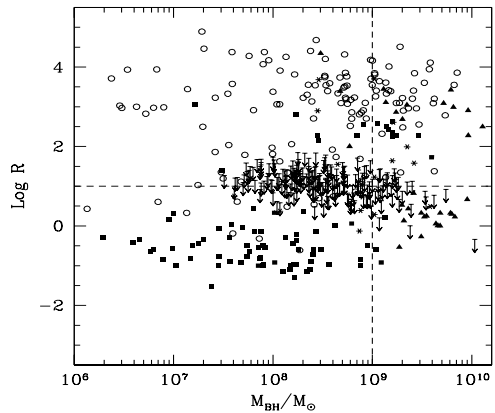,width=7cm,height=6cm,angle=0}
}
\caption{Radio loudness versus black hole mass for 452 AGN,
most at $z \simlt 1$.
There is no correlation between radio loudness and black hole mass.
In particular, at high mass ($M > 10^9 M\odot$) there
are similar distributions of black hole mass
for radio-loud ($R>10$) and radio-quiet AGN.
{\it Squares:} PG quasars, all at $z<0.5$;
{\it circles:} remaining AGN from Paper I, at $0<z<1$;
{\it triangles:} high-redshift quasars ($2<z<2.5$) from McIntosh et al. (1999);
{\it stars:} LBQS quasars, $0.5<z<1$;
{\it arrows:} upper limits for LBQS quasars, $0.5<z<1$.
Radio fluxes for AGN from the Parkes sample of Oshlack et al. (2002)
have been revised downward to account for relativistic beaming, following
Jarvis \& McLure (2002).
 }
\label{fig6}
\end{figure}

One of the most persistent and intriguing suggestions is that
radio loudness is closely tied to black hole mass. In particular, 
it has been suggested that radio luminosity correlates with 
black hole mass (Franceschini et al. 1998, but see Oshlack et al. 2002),
or that there might be a threshold effect such that radio loudness
requires a high black hole mass ($M > 10^9 M_\odot$; Laor 2000,
Lacy et al. 2001).
We find no such relation.
Figure~6 shows radio loudness versus black hole mass
for the largest compilation of AGN to date
(Woo \& Urry 2002b).

\section{Grand Unification of AGN and Galaxies}

It seems increasingly likely that AGN are normally evolving galaxies
going through a high-accretion-rate phase. This is supported by
several independent lines of evidence.
First, essentially all local normal galaxies 
(at least those with bulges) host supermassive black holes
(Kormendy \& Gebhardt 2001).
The correlation between black hole mass (most of which 
must be accumulated during the high-accretion period) 
and the host galaxy bulge luminosity and/or
stellar velocity dispersion directly implies 
at least a related evolution of AGN and galaxies.
Second, equating the integrated quasar light to accreted mass
(modulo some efficiency factor) requires episodic accretion
to occur for $\sim5-10$\% of the time in most galaxies in order
to avoid overly massive black holes (Cavaliere \& Padovani 1989).
Third, the host galaxies of AGN are indistinguishable from
normal galaxies, at least based on present imaging data
(Taylor et al. 1996; McLure et al. 1999; Urry et al. 2000;
Bettoni et al. 2001).
Finally, a few percent of Lyman-break galaxies have emission
lines, and appear to be some sort of AGN
(Steidel et al. 2002).

Suppose AGN and galaxies are precisely the same population,
with AGN simply representing a particular phase of normal galaxy evolution.
We call this the ``Grand Unification" hypothesis. 
The AGN phase in individual galaxies would occur when
epidodes of high accretion are triggered, 
and could be influenced by interactions
or mergers. As many have pointed out, this explains naturally why
there are more AGN at high redshift, where the galaxy density is far higher
and interactions are more common.

Obviously, for the better understanding of our Universe, it is
of great interest to test the Grand Unification hypothesis thoroughly. 
One key issue is whether AGN host galaxies are indeed 
indistinguishable from normal galaxies. Most such inferences
are based on imaging data; measuring the stellar velocity dispersions
in AGN host galaxies --- a project that several groups are now
undertaking --- is a critical next step.

In the presently preferred scenario, black hole accretion
coincides with bulge formation and starburst activity
(Kormendy \& Gebhardt 2001).
Jets may be generated in most or all galactic nuclei, with
a broad distribution of initial kinetic power
(cf. Meier 2001). Weak jets would be most prevalent by far,
and would lead to radio-quiet AGN, possibly with jet-initiated
winds and outflows, of the kind seen in many Seyfert galaxies.
More powerful jets would evolve into luminous radio sources once the 
dense gas associated with the starburst has cleared (Baker et al. 2002).
Testing this Grand Unified picture is the next step in
understanding AGN.

\bigskip

\acknowledgements

The topic of radio-loud AGN is enormous and it is impossible to cover the
territory with any completeness. My apologies for the many
omissions of significant work, and my gratitude to 
colleagues whose work forms the basis of this talk.
Fortunately, much of the detail was filled in by other
speakers at the excellent Paris AGN conference, whose
papers fill this volume.
I thank my students, Jong-Hak Woo and Matt O'Dowd, whose work on host
galaxies and black hole mass is reported here.
Finally, I thank Suzy Collin and Francoise Combes and
the other conference organizers who made our visit to
Paris such a productive and enjoyable one.


\begin{references}

\reference 
Antonucci, R., \& Miller, J. S. 1985, ApJ 297, 621

\reference 
Baker, J. C., Hunstead, R. W., Athreya, R. M., Barthel, P. D.,
de Silva, E., Lehnert, M. D., \& Saunders, R. D. E. 2002,
ApJ 568, 592

\reference 
Bettoni, D., Falomo, R., Fasano, G., Govoni, F., Salvo, M., \& 
Scarpa, R. 2001, A\&A, 380, 471

\reference 
Bicknell, G. V. 1995, ApJS, 101, 29

\reference 
Biretta, J. A., Zhou, F., \& Owen, F. N. 1995, ApJ 447, 582

\reference 
Blundell, K. M., \& Beasley, A. J. 1998, BAAS 30, 1418 (\#110.04)

\reference 
Bridle, A. H., \& Perley, R. A. 1984, ARAA 22, 319

\reference 
Brunthaler, A., et al. 2000, A\&A, 357, L45

\reference 
Cavaliere, A., \& Padovani, P. 1989, ApJ, 340, L5

\reference 
Celotti, A., \& Fabian, A. C. 1993, MNRAS 264, 228

\reference 
Celotti, A., Ghisellini, G., \& Chiaberge, M. 2001, MNRAS 321, L1

\reference 
Chiaberge, M., Capetti, A., \& Celotti, A. 2002, A\&A 394, 791

\reference 
De Young, D. S. 1993, ApJ Letters 405, L13 

\reference 
Falcke, H., Patnaik, A. R., \& Sherwood, W. 1996, ApJ 473, 13

\reference 
Falomo, R., Scarpa, R., Treves, A., \& Urry, C. M.
2000, ApJ 542, 731

\reference 
Falomo, R., Urry, C. M., Pesce, J. E., Scarpa, R., 
Treves, A., \& Giavalisco, M. 1997, ApJ 476, 113

\reference 
Franceschini, A., Vercellone, S., \& Fabian, A. C. 1998, MNRAS, 297, 817

\reference 
Garrington, S. T., \& Conway, R. G. 1991, MNRAS 250, 198

\reference 
Giovannini, G., Arbizzani, E., Feretti, L., Venturi, T., Cotton, W. D., 
Lara, L., \& Taylor, G. B. 1998, in {\it Radio Emission from Galactic 
and Extragalactic Compact Sources} (IAU Colloquium 164),
eds. J.A. Zensus, G.B. Taylor, \& J.M. Wrobel, 
(ASP Conf. Series), Vol. 144, p. 85.

\reference 
Jackson, C. A., \& Wall, J. V. 1999, MNRAS 304, 160

\reference 
Jarvis, M. J. \& McLure, R. J. 2002, MNRAS 337, 109

\reference 
Kaspi, S., et al. 2000, ApJ, 533, 631

\reference 
Kellermann, K. I., Sramek, R., Schmidt, M., Shaffer, D. B., \& Green, R.
1989, AJ, 98, 1195

\reference 
Kormendy, J., \& Gebhardt, K. 2001, in {\it 20th Texas Symposium on 
Relativistic Astrophysics}, eds. J. Craig Wheeler \& Hugo Martel,
(AIP Conf. Proc.) Vol. 586, p.363

\reference 
Laing, R. A. 1988, Nature 331, 149

\reference 
Laing, R. A., Jenkins, C. R., Wall, J. V., \& Unger, S. W. 1994, in
{\it The Physics of Active Galaxies}, ed. G. V. Bicknell, M. A. Dopita,
\& P. J. Quinn, (ASP Conf. Series), Vol. 54, p. 201

\reference 
Lacy, M., Ridgway, S., \& Trentham, N. 2001, ApJ Letters 551, 17

\reference 
Laor, A. 2000, ApJ, 543, L111

\reference 
Maraschi, L., \& Rovetti, F. 1994, ApJ 436, 79

\reference 
McIntosh, D. H., Rieke, M. J., Rix, H.-W., Foltz, C. B, 
\& Weymann, R. J. 1999, ApJ, 514, 40

\reference 
McLure, R. J., Kukula, M. J., Dunlop, J. S., Baum, S. A., \& 
O'Dea, C. P. 1999, MNRAS, 308, 377

\reference 
Meier, D. L. 2001, ApJ 548, L9

\reference 
Miyoshi, M., Moran, J., Herrnstein, J., Greenhill, L., Nakai, N., Diamond, P.,
\& Inoue, M. 1995, Nature 373, 127

\reference 
O'Dowd, M., Urry, C. M., \& Scarpa, R. 2002, ApJ, 580, 93

\reference 
Oshlack, A., Webster, R., \& Whiting, M. 2002, ApJ, 575, 810

\reference 
Owen, F. N., \& Ledlow, M. J. 1994, in {\it The Physics of
Active Galaxies}, ed. G. V. Bicknell, M. A. Dopita \& P. J. Quinn
(ASP Conf. Series), Vol. 54, p.31

\reference 
Sambruna, R. M., et al. 2002, ApJ 571, 206

\reference 
Scarpa, R., Urry, C. M., Falomo, R., Pesce, J. E., 
\& Treves, A. 2000, ApJ 532, 740

\reference 
Scarpa, R., Urry, C. M., Padovani, P., Calzetti, D., \& O'Dowd, M.
2001, ApJ 544, 258

\reference 
Steidel, C. C., Hunt, M. P., Shapley, A. E., Adelberger, K. L., 
Pettini, M., Dickinson , M., \& Giavalisco, M. 2002, ApJ 576 653

\reference 
Tavecchio, F., Maraschi, L., Sambruna, R. M., \& Urry, C. M. 
2000, ApJ Letters 544, L23

\reference 
Taylor, G. L., Dunlop, J. S., Hughes, D. H., \& Robson, E. I. 1996, MNRAS, 283, 930

\reference 
Tran, H. D. 2001, ApJ 554, L19

\reference 
Urry, C. M., Falomo, R., Scarpa, R., Pesce, J. E., Treves, A.,
\& Giavalisco, M. 1999, ApJ 512, 88

\reference 
Urry, C. M., \& Padovani, P. 1995, PASP 107, 803

\reference 
Urry, C. M., Scarpa, R., O'Dowd, M., Falomo, R., Pesce, J. E.,
\& Treves, A. 2000, ApJ 532, 816

\reference 
Vagnetti, F., Cavaliere, A., \& Giallongo, E. 1991, ApJ 368, 366

\reference 
White, R. L., et al. 2001, ApJS 126, 133

\reference 
Woo, J.-H., \& Urry, C. M. 2002a, ApJ 579, 530

\reference 
Woo, J.-H., \& Urry, C. M. 2002b, ApJ Letters 581, L5

\end{references}
\end{document}